\documentclass[copyright,creativecommons]{eptcs}

\usepackage{iftex}

\ifpdf
  \usepackage{underscore}         % Only needed if you use pdflatex.
  \usepackage[T1]{fontenc}        % Recommended with pdflatex
\else
  \usepackage{breakurl}           % Not needed if you use pdflatex only.
\fi

\usepackage[T1]{fontenc}
\usepackage{lmodern}
\usepackage{graphicx}
\usepackage{amsmath,amsfonts}
\usepackage{listings}
\usepackage{color}
\usepackage{cite}
\usepackage{adjustbox}
\usepackage{subcaption}

\lstdefinelanguage{civl}{
  morekeywords=[1]{$assume, $assert, $mem, $mem_disjoint, $mem_empty, $read_set_pop, $read_set_push, $write_set_pop, $write_set_push, $local_start, $local_end, $parfor, $input, $output},
  morekeywords=[2]{int, void, for, double, if, else, while},
  morekeywords=[3]{pragma, section, sections, shared, private, reduction, atomic, seq_cst, critical, parallel, omp, schedule, dynamic, static, guided, master, single, firstprivate, lastprivate, threadprivate, ordered, omp_set_num_threads, omp_get_num_threads, omp_get_thread_num, omp_init_lock, omp_destroy_lock, omp_set_lock, omp_unset_lock, omp_test_lock, omp_get_wtime, num_threads, barrier, read, write},
  alsoletter={$},
  sensitive=false,
  morecomment=[l]{//},
  morecomment=[s]{/*}{*/},
  keywordstyle=[1]\ttfamily\color[rgb]{0,.3,.7},
  keywordstyle=[2]\ttfamily\color[rgb]{0.133,0.545,0.133},
  keywordstyle=[3]\ttfamily\color[rgb]{0.545,0.133,0.133},
  commentstyle=\color[rgb]{0.5,0.5,0.5},
  stringstyle={\color[rgb]{0.75,0.49,0.07}},
  basicstyle=\ttfamily,
  columns=fullflexible,
  keepspaces=true,
  upquote=true,
  breaklines=true,
  breakatwhitespace=true,
  numbers=left,
  frame=L,
  xleftmargin=20pt
}

\lstdefinelanguage{civlsmall}{
  morekeywords=[1]{$assume, $assert, $mem, $mem_disjoint, $mem_empty, $read_set_pop, $read_set_push, $write_set_pop, $write_set_push, $local_start, $local_end, $parfor, $input, $output},
  morekeywords=[2]{int, void, for, double, if, else, while},
  morekeywords=[3]{pragma, section, sections, shared, private, reduction, atomic, seq_cst, critical, parallel, omp, schedule, dynamic, static, guided, master, single, firstprivate, lastprivate, threadprivate, ordered, omp_set_num_threads, omp_get_num_threads, omp_get_thread_num, omp_init_lock, omp_destroy_lock, omp_set_lock, omp_unset_lock, omp_test_lock, omp_get_wtime, num_threads, barrier, read, write},
  basicstyle=\ttfamily\scriptsize,
  columns=fullflexible,
  keepspaces=true,
  alsoletter={$},
  sensitive=false,
  morecomment=[l]{//},
  morecomment=[s]{/*}{*/},
  keywordstyle=[1]\ttfamily\color[rgb]{0,.3,.7},
  keywordstyle=[2]\ttfamily\color[rgb]{0.133,0.545,0.133},
  keywordstyle=[3]\ttfamily\color[rgb]{0.545,0.133,0.133},
  commentstyle=\color[rgb]{0.5,0.5,0.5},
  stringstyle={\color[rgb]{0.75,0.49,0.07}},
  upquote=true,
  breaklines=true,
  breakatwhitespace=true,
  numbers=left,
  frame=L,
  xleftmargin=20pt
}

\newcommand{\code}[1]{\texttt{#1}}

\title{Verifying a Sparse Matrix Algorithm Using Symbolic Execution}
\author{Alexander C. Wilton
\institute{Department of Computer and Information Sciences\\
University of Delaware, Newark, DE 19716, USA}
\email{awilton@udel.edu}
}
\def\titlerunning{Verifying a Sparse Matrix Algorithm Using Symbolic Execution}
\def\authorrunning{A.C. Wilton}
\begin{document}
\maketitle

\begin{abstract}
  Scientific software is, by its very nature, complex. It is
  mathematical and highly optimized which makes it prone to subtle
  bugs not as easily detected by traditional testing. We outline how
  symbolic execution can be used to write tests similar to traditional
  unit tests while providing stronger verification guarantees and
  apply this methodology to a sparse matrix algorithm.
\end{abstract}

\section{Introduction}

Scientific software has become ubiquitous across almost every field of
science due to continuous advancements in computing. This type of
software is usually designed to take on massive problems whose results
are often used in critical decisions, making its correctness
paramount.

However, this is not always easy. Scientific software is typically
very mathematical and highly optimized which leaves room for subtle
bugs that are not easily caught with traditional testing
\cite{kanewalaTestingScientificSoftware2014}.

Lightweight static analysis tools can be used but they are typically
too inaccurate to be useful in complex settings
\cite{johnsonWhyDontSoftware2013a}. Heavyweight approaches to
verification such as mechanized proofs can work in theory, but they
involve an enormous amount of specialized effort to create and
maintain \cite{leroyFormalVerificationRealistic2009}.

An intermediate approach to verification is through \textit{symbolic
  execution} \cite{kingSymbolicExecutionProgram1976,
  clarkeSystemGenerateTest1976, cadarKLEEUnassistedAutomatic,
  baldoniSurveySymbolicExecution2018}. This is a technique which
simulates execution of a program using symbolic expressions for its
values instead of concrete values. This allows for a potentially
infinite number of inputs to be reasoned about at once.

In this paper we outline how symbolic execution can be used in
scientific software similarly to unit testing while providing stronger
verification guarantees. We demonstrate this on a matrix-vector
multiplication algorithm adapted from \cite{kellisonLAProof2023} using
the CIVL model checker and symbolic execution tool
\cite{siegelCIVLConcurrencyIntermediate2015, civl:2025:web}. The
tricky part is that this algorithm works on matrices stored in the
widely used \textit{compressed row storage} (CRS) format, while the
vector is dense which means it is just stored as an array. The CRS
format is more commonly referred to as the \textit{compressed sparse
  row} (CSR) format. However, we opted to use this less common name to
match the language used in the paper \cite{kellisonLAProof2023} that
the multiplication algorithm is borrowed from.

In the rest of the paper we provide a brief overview of symbolic
execution and CIVL in section \ref{sec:CIVL}. Then we outline our
approach to verifying the CRS multiplication algorithm in Section
\ref{sec:verify} with an emphasis on the general approach to using
CIVL in this way. Finally we give concluding remarks in Section
\ref{sec:conclusion}.

\section{Symbolic Execution}\label{sec:CIVL}

Symbolic execution is a well known verification technique in which a
full simulation of the input program is executed with a ``symbolic''
semantics: each variable's value is represented by a symbolic
expression instead of a concrete value. This allows for representing
and reasoning over a potentially infinite number of inputs.

Branching and non-deterministic behavior is supported via backtracking
so that the entire space of reachable states is exhaustively searched
for errors. Each state includes a hidden boolean variable called the
\textit{path condition} which stores the set of assumptions and
branches taken in the current execution. If a particular execution
reaches a point in which the path condition is unsatisfiable, then the
branch is deemed \textit{infeasible} and is pruned from the search.

Because all feasible branches are checked, symbolic execution can
easily end up running indefinitely if there is some loop in the
program whose termination depends on a symbolic value. To remedy this,
such inputs need to be given small bounds by the user. In practice,
these bounds are usually placed on the size of some data structure
while the data this structure actually holds is left unbounded. While
this may appear to be a serious limitation, experience supports the
\emph{small scope hypothesis} \cite{jacksonElementsOfStyle1996,
  jackson:2019:alloy}. This posits that almost all bugs will appear on
inputs within some small bounds when a system is properly
parameterized.

The path condition is also used when checking assertions. If an
assertion is reached and it is determined that the current path
condition does not imply the assertion then an error is reported to
the user. To make these validity/unsatisfiability checks, external SMT
(Satisfiability Modulo Theories) solvers are often used.

\subsection{CIVL}
CIVL is a symbolic execution tool bundled with its own intermediate
language CIVL-C. It has a front-end that currently supports C and
FORTRAN programs as input. CIVL-C offers programmers a familiar syntax
and semantics because it is a large subset of standard C with
additional language features supporting concurrency, specification and
verification.

The SMT solvers CIVL currently supports are CVC4
\cite{barrettCVC42011} and Z3 \cite{demouraZ3EfficientSMT2008}. An
internal symbolic reasoner is also present which allows for many of
these external SMT queries to be simplified or avoided entirely.

CIVL is also a model checker \cite{clarkeModelCheckingState2012} which
naturally extends its symbolic execution framework to support
concurrent programs. This includes the use of several concurrency
dialects such as OpenMP or CUDA-C. However, for better clarity and
focus on symbolic execution, we will be restricting our attention to
sequential programs.

There are other symbolic execution tools
\cite{cadarKLEEUnassistedAutomatic, maDirectedSymbolicExecution2011,
  pasareanuSymbolicPathFinderSymbolic2010} which could just as easily
be used in the way we present in this paper. The reason for choosing
CIVL in this case simply comes down to the author's familiarity with
the tool and its direct support for C programs.

\subsubsection{Modeling Floating Point Numbers as Reals}

Floating point numbers in CIVL are modeled as real numbers. While
floating point properties are important, they add significant
complexity to both specification and verification. We argue that
specifications based on real numbers are often more appropriate as an
initial verification target.

The primary reason for this is that they are independent of
implementation specific details such as the order that floating point
operations are performed. This is what allows for the ``code as
specification'' approach highlighted in this paper. For example,
Strassen's algorithm is not bit-level equivalent to naive matrix
multiplication but it is equivalent using real numbers.

Additionally, violations of a specification based on reals represent
logical errors which are usually more pressing and common. After these
bugs are ironed out, another tool can be used to analyze floating
point properties if desired. Ideal real models of arithmetic are
complementary to approaches focused on floating points and generally
offer a quicker initial verification pass.

\section{Verifying a Multiplication Algorithm}\label{sec:verify}

\begin{figure}
  \centering
  \begin{minipage}[b][][b]{0.5\linewidth}
    \begin{lstlisting}[language=civlsmall]
#ifndef SPARSE_H
#define SPARSE_H

struct crs_matrix {
  double *val;
  unsigned *col_ind;
  unsigned *row_ptr;
  unsigned rows, cols;
};

void crs_matrix_vector_multiply(struct crs_matrix *m, double *v, double *p);

#endif\end{lstlisting}
    \caption{\code{sparse.h}}
    \label{fig:code:sparse-h}
  \end{minipage}%
  \begin{minipage}[b][][b]{0.5\linewidth}
    \begin{lstlisting}[language=civlsmall,escapechar=@]
#include "sparse.h"
void crs_matrix_vector_multiply (
       struct crs_matrix *m,
       double *v, double *p) {
  unsigned i, rows=m->rows;
  double *val = m->val;
  unsigned *col_ind = m->col_ind;
  unsigned *row_ptr = m->row_ptr;
  unsigned next=row_ptr[0];
  for (i=0; i<rows; i++) {
    double s=0.0;
    unsigned h=next; @\label{line:sparse-c:bug-1}@
    next=row_ptr[i+1]; @\label{line:sparse-c:bug-2}@
    for (; h<next; h++) {
      double x = val[h];
      unsigned j = col_ind[h];
      double y = v[j];
      s = x*y+s;
    }
    p[i]=s;
  }
}\end{lstlisting}
    \caption{\code{sparse.c} : \code{m} and \code{v} are input parameters and \code{p} is an output}
    \label{fig:code:sparse-c}
  \end{minipage}
\end{figure}

\begin{figure}
  \centering
  \begin{minipage}[b][][b]{0.5\linewidth}
    \begin{lstlisting}[language=civlsmall]
#ifndef MATRIX_H
#define MATRIX_H

typedef struct $mat {
  int n, m; // num rows, columns;
  double data[][];
} $mat;

void $mat_vec_mul($mat mat, double * v, double *p);

#endif\end{lstlisting}
    \caption{\code{matrix.cvh}}
    \label{fig:code:matrix-h}
  \end{minipage}%
  \begin{minipage}[b][][b]{0.5\linewidth}
    \begin{lstlisting}[language=civlsmall]
#include "matrix.cvh"
void $mat_vec_mul($mat mat, double * v, double *p) {
  int n = mat.n, m = mat.m;
  for (int i=0; i<n; i++) {
    double s = 0.0;
    for (int j=0; j<m; j++)
      s += mat.data[i][j]*v[j];
    p[i] = s;
  }
}\end{lstlisting}
    \caption{\code{matrix.cvl}}
    \label{fig:code:matrix-c}
  \end{minipage}
\end{figure}

Verifying algorithms with CIVL is often much like writing a
traditional unit test. The general workflow consists of writing a
program, called a \textit{driver}, which:
\begin{enumerate}
\item Generates inputs for the test;
\item Executes the algorithm being tested;
\item Calculates the expected result using some trusted source;
\item Compares the results;
\item Performs any tear-down/cleanup.
\end{enumerate}

Because scientific software is usually built upon a foundation of
mathematical libraries, there are many software components which
are amenable to this kind of verification.

To demonstrate this we will apply CIVL in this way to a multiplication
algorithm between a CRS matrix and a dense vector. It is extracted
from the source code located at \cite{LAProof:2025:web} for the paper
\cite{kellisonLAProof2023}. 

The CRS format for a matrix is a common way to efficiently represent
sparse matrices. The C structure used to represent a CRS matrix is
declared in a header file \code{sparse.h} shown in Figure
\ref{fig:code:sparse-h}. The fields \code{rows} and \code{cols} store
the number of rows and columns of the matrix. The field \code{val}
stores all the non-zero entries of the matrix as a single array in
row-order. Array \code{col\_ind} has the same length as \code{val} and
stores which column each corresponding entry of \code{val} is
located. The array \code{row\_ptr} has size $\code{rows} + 1$ and is
monotonically increasing. The $i$th entry of \code{row\_ptr} holds the
index of \code{val} and \code{col\_ind} where the $i$th row starts.

The multiplication algorithm we wish to verify is implemented in the
source file \code{sparse.c} shown in Figure
\ref{fig:code:sparse-c}. It takes as input a CRS matrix \code{m}, a
(dense) vector \code{v}, and a pointer \code{p} which will point to
the result of the multiplication when the call returns.

\subsection{Specification}
When CIVL executes code it will check for many different types of
errors such as dividing by zero or accessing an array out of
bounds. However we are also interested in the functional correctness
of the multiplication algorithm. This requires determining what it
means for the algorithm to be ``correct.''

We know what a matrix is as a mathematical concept and what it means to multiply it
with a vector. The CRS structure is a way to \emph{represent} a
matrix. So to say this multiplication algorithm is correct means that
the result of executing the function \code{crs\_matrix\_vector\_multiply}
is the same as performing the mathematical operation on
the standard matrix that the input CRS structure represents.

So the key to specifying any algorithm involving a CRS structure is to
describe exactly how such a structure represents a standard
matrix. This is done by defining a \emph{representation
  function}.

In the paper \cite{kellisonLAProof2023} that this example is taken
from, a representation relation is used in a similar way. The
difference is that in \cite{kellisonLAProof2023}, a declarative
approach is used, whereas here we use executable code to make this
correspondence.

We created a header file \code{matrix.cvh} shown in Figure
\ref{fig:code:matrix-h} that contains the CIVL-C structure which
represents a standard matrix. The multiplication function for this
matrix type is in the source file \code{matrix.cvl} shown in Figure
\ref{fig:code:matrix-c}.

The representation function for a CRS matrix is implemented on lines
\ref{line:driver:rep-start}--\ref{line:driver:rep-end} of our driver
\code{driver.cvl} presented in Figure \ref{fig:code:driver}. We kept
this function in the driver for simplicity but in practice it may be
better to separate this out into its own library for specifying and
verifying algorithms related to CRS matrices.

\begin{figure}
  \centering
  \begin{lstlisting}[language=civlsmall,escapechar=@,xleftmargin=.2\textwidth]
#include <stdlib.h>
#include <pointer.cvh>
#include "sparse.h"
#include "matrix.cvh"

$input unsigned N_B = 3, M_B = 3, N, M; @\label{line:driver:input-start}@
$assume (1<=N && N<=N_B && 1<=M && M<=M_B); @\label{line:driver:input-assume}@
$input double V[M], A[N*M]; @\label{line:driver:input-pool}@

/* Fills in p[0],...,p[len-1] with a strictly increasing sequence
   of integers in [0,max].  Precondition: 0 <= len <= max+1 */
void strict_inc(unsigned * p, unsigned len, unsigned max) {
  for (int i=0; i<len; i++) {
    unsigned a = (i == 0 ? 0 : p[i-1]+1), b = max - len + i + 1;
    p[i] = a + $choose_int(b-a+1); // choose in a..b
  }
}

struct crs_matrix make_crs(unsigned n, unsigned m) { @\label{line:driver:make-crs-start}@
  unsigned * row_ptr = malloc((n+1)*sizeof(unsigned)); @\label{line:driver:make-crs-row-malloc}@
  row_ptr[0] = 0;
  for (int i=1; i<=n; i++) @\label{line:driver:row-ptr-1}@
    row_ptr[i] = row_ptr[i-1]+$choose_int(m+1); @\label{line:driver:row-ptr-2}@
  unsigned NZ = row_ptr[n];
  unsigned * col_ind = malloc(NZ*sizeof(unsigned));
  for (int i=0; i<n; i++)
    strict_inc(col_ind+row_ptr[i], row_ptr[i+1]-row_ptr[i], m-1);
  double * val = malloc(NZ*sizeof(double));
  for (int i=0; i<NZ; i++) val[i] = A[i];
  return (struct crs_matrix){ val, col_ind, row_ptr, n, m };
} @\label{line:driver:input-end}@

void destroy_crs(struct crs_matrix mat) { @\label{line:driver:cleanup-start}@
  free(mat.val);
  free(mat.col_ind);
  free(mat.row_ptr);
} @\label{line:driver:cleanup-end}@

$mat $mat_crs(struct crs_matrix crs) { @\label{line:driver:rep-start}@
  unsigned n = crs.rows, m = crs.cols;
  $mat mat;
  mat.n = n;
  mat.m = m;
  mat.data = (double[n][m])$lambda(int i,j) 0.0;
  for (int i=0; i<n; i++) {
    unsigned r = crs.row_ptr[i],
      rnxt = crs.row_ptr[i+1];
    for (int k=r; k<rnxt; k++)
      mat.data[i][crs.col_ind[k]] = crs.val[k];
  }
  return mat;
} @\label{line:driver:rep-end}@

int main() { @\label{line:driver:main-start}@
  double v[M], actual[N], expected[N];
  for (int i=0; i<M; i++) v[i] = V[i]; @\label{line:driver:v-init}@
  struct crs_matrix mat = make_crs(N, M); @\label{line:driver:make-crs}@
  crs_matrix_vector_multiply(&mat, v, actual);
  $mat dense = $mat_crs(mat);
  $mat_vec_mul(dense, v, expected); @\label{line:driver:dense-mul}@
  $assert($equals(actual, expected)); @\label{line:driver:assert}@
  destroy_crs(mat);
} @\label{line:driver:main-end}@\end{lstlisting}
    \caption{\code{driver.cvl}}
    \label{fig:code:driver}
  \end{figure}
  
\subsection{Generating Inputs}

Just like a unit test, our driver file must generate inputs for the
algorithm. Lines
\ref{line:driver:input-start}--\ref{line:driver:input-end} of our
driver presented in Figure \ref{fig:code:driver} are responsible for
this.

CIVL-C provides a type qualifier \code{\$input} for global variables
which marks them as read-only and initializes them with an arbitrary
value of their type. This is used on lines \ref{line:driver:input-start} and
\ref{line:driver:input-pool}.

The variables \code{N} and \code{M} represent the number of rows and
columns, respectively. These need to be bounded because otherwise the
state space will be infinite since we usually loop over these
variables.

CIVL allows for \code{\$input} variables to be given specific concrete
values from the command line. This could be used for verification with
a specific number of rows and columns. However it is more convenient
to check a range of values for \code{N} and \code{M}. So instead we
add two additional \code{\$input} variables \code{N\_B} and
\code{M\_B} which are used to provide an upper bound on the values of
\code{N} and \code{M} using the \code{\$assume} statement on line
\ref{line:driver:input-assume}. By default these bounds are
initialized to $3$ but can be overridden from the command line.

The \code{\$input} array \code{V} is used by the \code{main} function
of our driver to fill out an array that represents our input
vector. Generating the CRS matrix itself is trickier.

It is possible to use \code{\$input} on a global variable of type
\code{struct crs\_matrix} but there is a problem with this. Not every
possible object of such a struct actually describes a valid CRS
matrix. For instance \code{val} and \code{col\_ind} must both point to
arrays with the same length. The sequence \code{row\_ptr[0]},
\code{row\_ptr[1]}, $\ldots$ must be monotonically increasing. These
are examples of data structure invariants that many functions which
consume CRS structures require of their inputs.

We could describe these invariants using the \code{\$assume} statement
provided by CIVL-C. This offsets a lot of complex reasoning to the SMT
solvers which can result in unsolved queries causing spurious reports
of failure. Alternatively we can non-deterministically construct an
arbitrary (valid) input with code similar to how a unit test might
randomly generate its inputs. This is usually more intuitive for
programmers inexperienced with formal logic, but it can blow up the
state space due to an increased number of branches. We take the latter
approach. This is implemented by the function \code{make\_crs} seen on
lines \ref{line:driver:make-crs-start}--\ref{line:driver:input-end}
which we will briefly explain.

After allocating space for \code{row\_ptr} on line
\ref{line:driver:make-crs-row-malloc}, it is filled with monotonically
increasing data representing each row's starting index in \code{val}
and \code{col\_ind} (lines
\ref{line:driver:row-ptr-1}--\ref{line:driver:row-ptr-2}). This is done
using CIVL-C's \code{\$choose\_int(int n)} expression which
non-deterministically returns an integer value between 0 (inclusive)
and \code{n} (exclusive). A similar process is used to fill out the
values of \code{col\_ind} using a helper function
\code{strict\_inc}. Finally, \code{val} is filled out using the
\code{\$input} array \code{A} declared earlier.

Under symbolic execution, when this function returns, the actual
entries of the matrix will all be symbolic. The values of
\code{row\_ptr} and \code{col\_ind} will contain concrete values
chosen non-deterministically. To get a sense for this, we inserted
print statements (not shown) at the end of \code{make\_crs}. Because
all different choices get explored in symbolic execution, this results
in all explored inputs to be printed. Here is a small snippet of this
output:

\begin{lstlisting}[basicstyle=\ttfamily\scriptsize,numbers=none]
--------
n: 2 m: 3
val: [ X_A[0] X_A[1] X_A[2] ]
col_ind: [ 0 1 2 ]
row_ptr: [ 0 0 3 ]
--------

--------
n: 1 m: 3
val: [ ]
col_ind: [ ]
row_ptr: [ 0 0 ]
--------

--------
n: 3 m: 3
val: [ X_A[0] ]
col_ind: [ 0 ]
row_ptr: [ 0 1 1 1 ]
--------
\end{lstlisting}

A value of the form \code{X\_A[i]} represents the
$\code{i}^{\text{th}}$ index into the symbolic constant \code{X\_A}
assigned to the global \code{\$input} variable \code{A} on
initialization. Since this array is unconstrained, each of these
symbolic values represent an arbitrary real number. Therefore, if CIVL
reports the algorithm is bug-free then the program is correct for all
possible CRS matrices with up to \code{N\_B} rows and \code{M\_B}
columns.

\subsection{The Driver}

The \code{main} function of our driver can be seen on lines
\ref{line:driver:main-start}--\ref{line:driver:main-end} of Figure
\ref{fig:code:driver}. The array variable \code{v} represents our
input vector and is initialized as such with the values of our global
\code{\$input} variable \code{V} on line \ref{line:driver:v-init}. The
function \code{make\_crs} used to generate our input CRS matrix is
called on line \ref{line:driver:make-crs}. It initializes a
\code{crs\_matrix} variable \code{mat}.

With the inputs generated, we call
\code{crs\_matrix\_vector\_multiply} using the array \code{actual} to
store the results. Then, as described earlier, we create the matrix
\code{dense} which our CRS matrix \code{mat} represents. We use it to
calculate the expected result with the function \code{\$mat\_vec\_mul}
on line \ref{line:driver:dense-mul}. The result of this call is stored
in the array \code{expected}.

Next we assert that the two results stored in \code{actual} and
\code{expected} are in fact equal on line \ref{line:driver:assert}. For
convenience, we use a built-in CIVL-C primitive \code{\$equals} which
performs a deep equality between two arrays.

Finally we free any allocated memory made for the test. Cleanup is
important in the context of verification because memory leaks are
checked by CIVL. In this case, cleanup simply involves freeing the
memory allocated by \code{make\_crs}. A simple helper method
\code{destroy\_crs} is provided on lines
\ref{line:driver:cleanup-start}--\ref{line:driver:cleanup-end} which
does this.

\subsection{Output}\label{subsec:output}

To run the driver we use the command
\begin{verbatim}
     civl verify driver.cvl matrix.cvl sparse.c
\end{verbatim}
This links the three source files together effectively into one
single program for CIVL to analyze. Running this command results in
the following output:
\begin{lstlisting}[basicstyle=\ttfamily\scriptsize,numbers=none]
=== Source files ===
driver.cvl  (driver.cvl)
sparse.h  (sparse.h)
matrix.cvh  (matrix.cvh)
matrix.cvl  (matrix.cvl)
sparse.c  (sparse.c)

=== Command ===
civl verify driver.cvl matrix.cvl sparse.c 

=== Stats ===
   time (s)          : 9.36        transitions  : 108128        
   memory (bytes)    : 4.194304E8  trace steps  : 78911         
   max process count : 1           valid calls  : 281776        
   states            : 78239       provers      : cvc4, z3, why3
   states saved      : 113029      prover calls : 19            
   state matches     : 673         

=== Result ===
All errors marked with '+' are absent on all executions.
 + Dereference errors                  + Functional equivalence violations
 + Internal errors                     + Library loading errors
 + Other errors                        + Assertion violations
 + Communication errors                + Writes to constant variables
 + Absolute deadlocks                  + Division by zero
 + Writes to $input variables          + Invalid casts
 + Malloc errors                       + Memory leaks
 + Memory management errors            + MPI usage errors
 + Out of bounds errors                + Reads from $output variables
 + Pointer errors                      + Process leaks
 + Sequence errors                       Non-termination
 + Use of undefined values             + Union errors
\end{lstlisting}

Using a 2020 MacBook Air with a 1.2 GHz Quad-Core Intel Core i7, CIVL
was able to verify this program is free of any of the checked errors
for all CRS matrices with up to three rows and three columns in under
ten seconds. Note that non-termination is not checked by default
since this is not always in error in some programs.

We can push CIVL further by increasing the bounds of our matrix from
the command line. Recall that the \code{\$input} variable \code{M\_B}
bounded the number of columns of our matrix. We can rerun the example
while setting \code{M\_B} to a value of $4$ with the command
\begin{verbatim}
     civl verify -inputM_B=4 driver.cvl matrix.cvl sparse.c
\end{verbatim}
CIVL is still able to succeed but it takes around fifty seconds to do
so because the number of states start to explode.

We can see how CIVL works when there is a bug by intentionally
introducing one. One easy-to-make mistake in this algorithm would be
to accidentally switch the order of lines \ref{line:sparse-c:bug-1}
and \ref{line:sparse-c:bug-2} in the multiplication algorithm (Figure
\ref{fig:code:sparse-c}). Running CIVL on this produces the following
output:
\begin{lstlisting}[basicstyle=\ttfamily\footnotesize]
Violation 0 encountered at depth 113:
CIVL execution violation in p0
(property: ASSERTION_VIOLATION, certainty: PROVEABLE) at
$assert($equals(actual, expected))
driver.cvl:61:2-35 | $assert($equals(actual, expected))
\end{lstlisting}

Additional information is printed which we have omitted for
space. When an error is found, CIVL has several capabilities to assist
in pinpointing what went wrong. States and transitions taken can be
printed in full detail and CIVL can attempt to find a minimal
execution for the error. Another useful tactic is to insert printing
statements into the code and then use the \code{replay} command which
re-executes the path the program took when reaching the error.

\section{Conclusion}\label{sec:conclusion}

Scientific software often contains many highly optimized
components for which unit testing does not suffice. We argue that a
symbolic execution tool like CIVL can remedy this. We make this
argument by outlining how to write a verification driver similarly to
a unit test but with the ability to provide stronger verification
guarantees. We apply this methodology to an example from the paper
\cite{kellisonLAProof2023} and show that it was effective in verifying
it for inputs within a small set of bounds.

While the techniques here provide much stronger correctness than
traditional testing, the approach taken in the original paper is much
more comprehensive. It provides an analysis of floating-point
precision and constructs a formal proof. However the two approaches
are not mutually exclusive. Using symbolic execution is a more
lightweight approach that can be applied to quickly find bugs, even as
the code is being developed. Once symbolic execution no longer finds
any bugs, a more elaborate deductive approach can then be taken.

\subsection*{Acknowledgments.}
This work was supported by U.S.\ National Science Foundation grant CCF-1955852.

\bibliographystyle{eptcs}
\bibliography{vss}

\begin{thebibliography}{10}
\providecommand{\bibitemdeclare}[2]{}
\providecommand{\surnamestart}{}
\providecommand{\surnameend}{}
\providecommand{\urlprefix}{Available at }
\providecommand{\url}[1]{\texttt{#1}}
\providecommand{\href}[2]{\texttt{#2}}
\providecommand{\urlalt}[2]{\href{#1}{#2}}
\providecommand{\doi}[1]{doi:\urlalt{https://doi.org/#1}{#1}}
\providecommand{\eprint}[1]{arXiv:\urlalt{https://arxiv.org/abs/#1}{#1}}
\providecommand{\bibinfo}[2]{#2}

\bibitemdeclare{article}{baldoniSurveySymbolicExecution2018}
\bibitem{baldoniSurveySymbolicExecution2018}
\bibinfo{author}{Roberto \surnamestart Baldoni\surnameend},
  \bibinfo{author}{Emilio \surnamestart Coppa\surnameend},
  \bibinfo{author}{Daniele~Cono \surnamestart D'elia\surnameend},
  \bibinfo{author}{Camil \surnamestart Demetrescu\surnameend} \&
  \bibinfo{author}{Irene \surnamestart Finocchi\surnameend}
  (\bibinfo{year}{2018}): \emph{\bibinfo{title}{A {{Survey}} of {{Symbolic
  Execution Techniques}}}}.
\newblock {\slshape \bibinfo{journal}{ACM Computing Surveys}}
  \bibinfo{volume}{51}(\bibinfo{number}{3}), pp. \bibinfo{pages}{50:1--50:39},
  \doi{10.1145/3182657}.

\bibitemdeclare{inproceedings}{barrettCVC42011}
\bibitem{barrettCVC42011}
\bibinfo{author}{Clark \surnamestart Barrett\surnameend},
  \bibinfo{author}{Christopher~L. \surnamestart Conway\surnameend},
  \bibinfo{author}{Morgan \surnamestart Deters\surnameend},
  \bibinfo{author}{Liana \surnamestart Hadarean\surnameend},
  \bibinfo{author}{Dejan \surnamestart Jovanovi{\'c}\surnameend},
  \bibinfo{author}{Tim \surnamestart King\surnameend}, \bibinfo{author}{Andrew
  \surnamestart Reynolds\surnameend} \& \bibinfo{author}{Cesare \surnamestart
  Tinelli\surnameend} (\bibinfo{year}{2011}): \emph{\bibinfo{title}{{{CVC4}}}}.
\newblock In \bibinfo{editor}{Ganesh \surnamestart Gopalakrishnan\surnameend}
  \& \bibinfo{editor}{Shaz \surnamestart Qadeer\surnameend}, editors: {\slshape
  \bibinfo{booktitle}{Computer {{Aided Verification}}}},
  \bibinfo{publisher}{Springer}, \bibinfo{address}{Berlin, Heidelberg}, pp.
  \bibinfo{pages}{171--177}, \doi{10.1007/978-3-642-22110-1_14}.

\bibitemdeclare{inproceedings}{cadarKLEEUnassistedAutomatic}
\bibitem{cadarKLEEUnassistedAutomatic}
\bibinfo{author}{Cristian \surnamestart Cadar\surnameend},
  \bibinfo{author}{Daniel \surnamestart Dunbar\surnameend} \&
  \bibinfo{author}{Dawson \surnamestart Engler\surnameend}
  (\bibinfo{year}{2008}): \emph{\bibinfo{title}{KLEE: unassisted and automatic
  generation of high-coverage tests for complex systems programs}}.
\newblock In: {\slshape \bibinfo{booktitle}{Proceedings of the 8th USENIX
  Conference on Operating Systems Design and Implementation}},
  \bibinfo{series}{OSDI'08}, \bibinfo{publisher}{USENIX Association},
  \bibinfo{address}{USA}, p. \bibinfo{pages}{209–224}.

\bibitemdeclare{misc}{civl:2025:web}
\bibitem{civl:2025:web}
\emph{\bibinfo{title}{{CIVL}: The {C}oncurrency {I}ntermediate {V}erification
  {L}anguage}}.
\newblock \urlprefix\url{https://civl.dev}.
\newblock \bibinfo{note}{Accessed 1 Feb 2025}.

\bibitemdeclare{incollection}{clarkeModelCheckingState2012}
\bibitem{clarkeModelCheckingState2012}
\bibinfo{author}{Edmund~M. \surnamestart Clarke\surnameend},
  \bibinfo{author}{William \surnamestart Klieber\surnameend},
  \bibinfo{author}{Milo{\v s} \surnamestart Nov{\'a}{\v c}ek\surnameend} \&
  \bibinfo{author}{Paolo \surnamestart Zuliani\surnameend}
  (\bibinfo{year}{2012}): \emph{\bibinfo{title}{Model {{Checking}} and the
  {{State Explosion Problem}}}}.
\newblock In \bibinfo{editor}{Bertrand \surnamestart Meyer\surnameend} \&
  \bibinfo{editor}{Martin \surnamestart Nordio\surnameend}, editors: {\slshape
  \bibinfo{booktitle}{Tools for {{Practical Software Verification}}: {{LASER}},
  {{International Summer School}} 2011, {{Elba Island}}, {{Italy}}, {{Revised
  Tutorial Lectures}}}}, \bibinfo{series}{Lecture {{Notes}} in {{Computer
  Science}}}, \bibinfo{publisher}{Springer}, \bibinfo{address}{Berlin,
  Heidelberg}, pp. \bibinfo{pages}{1--30}, \doi{10.1007/978-3-642-35746-6_1}.

\bibitemdeclare{article}{clarkeSystemGenerateTest1976}
\bibitem{clarkeSystemGenerateTest1976}
\bibinfo{author}{L.A. \surnamestart Clarke\surnameend} (\bibinfo{year}{1976}):
  \emph{\bibinfo{title}{A {{System}} to {{Generate Test Data}} and
  {{Symbolically Execute Programs}}}}.
\newblock {\slshape \bibinfo{journal}{IEEE Transactions on Software
  Engineering}} \bibinfo{volume}{SE-2}(\bibinfo{number}{3}), pp.
  \bibinfo{pages}{215--222}, \doi{10.1109/TSE.1976.233817}.

\bibitemdeclare{inproceedings}{demouraZ3EfficientSMT2008}
\bibitem{demouraZ3EfficientSMT2008}
\bibinfo{author}{Leonardo \surnamestart {de Moura}\surnameend} \&
  \bibinfo{author}{Nikolaj \surnamestart Bj{\o}rner\surnameend}
  (\bibinfo{year}{2008}): \emph{\bibinfo{title}{Z3: {{An Efficient SMT
  Solver}}}}.
\newblock In \bibinfo{editor}{C.~R. \surnamestart Ramakrishnan\surnameend} \&
  \bibinfo{editor}{Jakob \surnamestart Rehof\surnameend}, editors: {\slshape
  \bibinfo{booktitle}{Tools and {{Algorithms}} for the {{Construction}} and
  {{Analysis}} of {{Systems}}}}, \bibinfo{publisher}{Springer},
  \bibinfo{address}{Berlin, Heidelberg}, pp. \bibinfo{pages}{337--340},
  \doi{10.1007/978-3-540-78800-3_24}.

\bibitemdeclare{article}{jackson:2019:alloy}
\bibitem{jackson:2019:alloy}
\bibinfo{author}{Daniel \surnamestart Jackson\surnameend}
  (\bibinfo{year}{2019}): \emph{\bibinfo{title}{{A}lloy: a language and tool
  for exploring software designs}}.
\newblock {\slshape \bibinfo{journal}{Commun. ACM}}
  \bibinfo{volume}{62}(\bibinfo{number}{9}), pp. \bibinfo{pages}{66--76},
  \doi{10.1145/3338843}.

\bibitemdeclare{article}{jacksonElementsOfStyle1996}
\bibitem{jacksonElementsOfStyle1996}
\bibinfo{author}{Daniel \surnamestart Jackson\surnameend} \&
  \bibinfo{author}{Craig~A. \surnamestart Damon\surnameend}
  (\bibinfo{year}{1996}): \emph{\bibinfo{title}{Elements of style: analyzing a
  software design feature with a counterexample detector}}.
\newblock {\slshape \bibinfo{journal}{SIGSOFT Softw. Eng. Notes}}
  \bibinfo{volume}{21}(\bibinfo{number}{3}), p. \bibinfo{pages}{239–249},
  \doi{10.1145/226295.226322}.

\bibitemdeclare{inproceedings}{johnsonWhyDontSoftware2013a}
\bibitem{johnsonWhyDontSoftware2013a}
\bibinfo{author}{Brittany \surnamestart Johnson\surnameend},
  \bibinfo{author}{Yoonki \surnamestart Song\surnameend},
  \bibinfo{author}{Emerson \surnamestart {Murphy-Hill}\surnameend} \&
  \bibinfo{author}{Robert \surnamestart Bowdidge\surnameend}
  (\bibinfo{year}{2013}): \emph{\bibinfo{title}{Why Don't Software Developers
  Use Static Analysis Tools to Find Bugs?}}
\newblock In: {\slshape \bibinfo{booktitle}{2013 35th {{International
  Conference}} on {{Software Engineering}} ({{ICSE}})}}, pp.
  \bibinfo{pages}{672--681}, \doi{10.1109/ICSE.2013.6606613}.

\bibitemdeclare{article}{kanewalaTestingScientificSoftware2014}
\bibitem{kanewalaTestingScientificSoftware2014}
\bibinfo{author}{Upulee \surnamestart Kanewala\surnameend} \&
  \bibinfo{author}{James~M. \surnamestart Bieman\surnameend}
  (\bibinfo{year}{2014}): \emph{\bibinfo{title}{Testing Scientific Software:
  {{A}} Systematic Literature Review}}.
\newblock {\slshape \bibinfo{journal}{Information and Software Technology}}
  \bibinfo{volume}{56}(\bibinfo{number}{10}), pp. \bibinfo{pages}{1219--1232},
  \doi{10.1016/j.infsof.2014.05.006}.

\bibitemdeclare{inproceedings}{kellisonLAProof2023}
\bibitem{kellisonLAProof2023}
\bibinfo{author}{Ariel~E. \surnamestart Kellison\surnameend},
  \bibinfo{author}{Andrew~W. \surnamestart Appel\surnameend},
  \bibinfo{author}{Mohit \surnamestart Tekriwal\surnameend} \&
  \bibinfo{author}{David \surnamestart Bindel\surnameend}
  (\bibinfo{year}{2023}): \emph{\bibinfo{title}{LAProof: A Library of Formal
  Proofs of Accuracy and Correctness for Linear Algebra Programs}}.
\newblock In: {\slshape \bibinfo{booktitle}{2023 IEEE 30th Symposium on
  Computer Arithmetic (ARITH)}}, \bibinfo{publisher}{IEEE}, pp.
  \bibinfo{pages}{36--43}, \doi{10.1109/ARITH58626.2023.00021}.

\bibitemdeclare{article}{kingSymbolicExecutionProgram1976}
\bibitem{kingSymbolicExecutionProgram1976}
\bibinfo{author}{James~C. \surnamestart King\surnameend}
  (\bibinfo{year}{1976}): \emph{\bibinfo{title}{Symbolic Execution and Program
  Testing}}.
\newblock {\slshape \bibinfo{journal}{Communications of the ACM}}
  \bibinfo{volume}{19}(\bibinfo{number}{7}), pp. \bibinfo{pages}{385--394},
  \doi{10.1145/360248.360252}.

\bibitemdeclare{misc}{LAProof:2025:web}
\bibitem{LAProof:2025:web}
\emph{\bibinfo{title}{LAProof}}.
\newblock \urlprefix\url{https://github.com/VeriNum/LAProof/tree/main}.
\newblock \bibinfo{note}{Accessed 1 Feb 2025}.

\bibitemdeclare{article}{leroyFormalVerificationRealistic2009}
\bibitem{leroyFormalVerificationRealistic2009}
\bibinfo{author}{Xavier \surnamestart Leroy\surnameend} (\bibinfo{year}{2009}):
  \emph{\bibinfo{title}{Formal Verification of a Realistic Compiler}}.
\newblock {\slshape \bibinfo{journal}{Commun. ACM}}
  \bibinfo{volume}{52}(\bibinfo{number}{7}), pp. \bibinfo{pages}{107--115},
  \doi{10.1145/1538788.1538814}.

\bibitemdeclare{inproceedings}{maDirectedSymbolicExecution2011}
\bibitem{maDirectedSymbolicExecution2011}
\bibinfo{author}{Kin-Keung \surnamestart Ma\surnameend}, \bibinfo{author}{Khoo
  \surnamestart Yit~Phang\surnameend}, \bibinfo{author}{Jeffrey~S.
  \surnamestart Foster\surnameend} \& \bibinfo{author}{Michael \surnamestart
  Hicks\surnameend} (\bibinfo{year}{2011}): \emph{\bibinfo{title}{Directed
  {{Symbolic Execution}}}}.
\newblock In \bibinfo{editor}{Eran \surnamestart Yahav\surnameend}, editor:
  {\slshape \bibinfo{booktitle}{Static {{Analysis}}}},
  \bibinfo{publisher}{Springer}, \bibinfo{address}{Berlin, Heidelberg}, pp.
  \bibinfo{pages}{95--111}, \doi{10.1007/978-3-642-23702-7_11}.

\bibitemdeclare{inproceedings}{pasareanuSymbolicPathFinderSymbolic2010}
\bibitem{pasareanuSymbolicPathFinderSymbolic2010}
\bibinfo{author}{Corina~S. \surnamestart P{\u a}s{\u a}reanu\surnameend} \&
  \bibinfo{author}{Neha \surnamestart Rungta\surnameend}
  (\bibinfo{year}{2010}): \emph{\bibinfo{title}{Symbolic {{PathFinder}}:
  Symbolic Execution of {{Java}} Bytecode}}.
\newblock In: {\slshape \bibinfo{booktitle}{Proceedings of the 25th
  {{IEEE}}/{{ACM International Conference}} on {{Automated Software
  Engineering}}}}, \bibinfo{series}{{{ASE}} '10},
  \bibinfo{publisher}{Association for Computing Machinery},
  \bibinfo{address}{New York, NY, USA}, pp. \bibinfo{pages}{179--180},
  \doi{10.1145/1858996.1859035}.

\bibitemdeclare{inproceedings}{siegelCIVLConcurrencyIntermediate2015}
\bibitem{siegelCIVLConcurrencyIntermediate2015}
\bibinfo{author}{Stephen~F. \surnamestart Siegel\surnameend},
  \bibinfo{author}{Manchun \surnamestart Zheng\surnameend},
  \bibinfo{author}{Ziqing \surnamestart Luo\surnameend},
  \bibinfo{author}{Timothy~K. \surnamestart Zirkel\surnameend},
  \bibinfo{author}{Andre~V. \surnamestart Marianiello\surnameend},
  \bibinfo{author}{John~G. \surnamestart Edenhofner\surnameend},
  \bibinfo{author}{Matthew~B. \surnamestart Dwyer\surnameend} \&
  \bibinfo{author}{Michael~S. \surnamestart Rogers\surnameend}
  (\bibinfo{year}{2015}): \emph{\bibinfo{title}{{{CIVL}}: The Concurrency
  Intermediate Verification Language}}.
\newblock In: {\slshape \bibinfo{booktitle}{{{SC}} '15: {{Proceedings}} of the
  {{International Conference}} for {{High Performance Computing}},
  {{Networking}}, {{Storage}} and {{Analysis}}}}, pp. \bibinfo{pages}{1--12},
  \doi{10.1145/2807591.2807635}.

\end{thebibliography}
\end{document}